\documentclass[prl,twocolumn,showpacs,amssymb]{revtex4}

\usepackage{psfrag,graphicx}
\usepackage{tikz}
\usepackage{pgfplots}
\usepackage{dcolumn}
\usepackage{bm}
\usepackage{amsfonts,amssymb,amsmath} 
\usepackage{xcolor}

\newcommand{\be}{\begin{equation}}
\newcommand{\ee}{\end{equation}}
\newcommand{\bq}{\begin{eqnarray}}
\newcommand{\eq}{\end{eqnarray}}

\newcommand{\SIGMA}{\mbox{\boldmath${\sigma}$}}

\newcommand{\jj}{\boldsymbol{j}}
\newcommand{\sv}{\boldsymbol{s}}
\newcommand{\pp}{\boldsymbol{p}}
\newcommand{\zz}{\boldsymbol{z}}

\begin{document}

\title{Induced topological order at the boundary of 3D topological superconductors } 

\author{Peter Finch}
\affiliation{School of Physics and Astronomy, University of Leeds, Leeds, LS2 9JT, United Kingdom}
\author{James de Lisle}
\affiliation{School of Physics and Astronomy, University of Leeds, Leeds, LS2 9JT, United Kingdom}
\author{Giandomenico Palumbo}
\affiliation{School of Physics and Astronomy, University of Leeds, Leeds, LS2 9JT, United Kingdom}
\author{Jiannis K. Pachos}
\affiliation{School of Physics and Astronomy, University of Leeds, Leeds, LS2 9JT, United Kingdom}

\date{\today}

\pacs{73.20.At, 74.20.Rp, 03.65.Vf, 71.10.Pm, 74.90.+n}

\begin{abstract}

\noindent
We present tight-binding models of 3D topological superconductors in class DIII that support a variety of winding numbers. We show that gapless Majorana surface states emerge at their boundary in agreement with the bulk-boundary correspondence. At the presence of a Zeeman field the surface states become gapped and the boundary behaves as a 2D superconductor in class D. Importantly, the 2D and 3D winding numbers are in agreement signifying that the topological order of the boundary is induced by the order of the 3D bulk. Hence, the boundary of a 3D topological superconductor in class DIII can be used for the robust realisation of localised Majorana zero modes.

\end{abstract}

\maketitle

{\bf \em Introduction:--} 1D and 2D tight-binding topological superconductors (TS) are a commonly employed medium to probe localised Majorana zero modes (Majoranas) with non-Abelian statistics. 1D tight-binding TS are employed to theoretically investigate the properties of Majoranas and experimentally verify their existence~\cite{KitaevWire}. 2D tight-binding TS provide exactly solvable models, where Majoranas exhibit rich behaviour~\cite{KitaevHoney,Ville1,Ville2}. Despite the success of these models little work has been done in relation to 3D tight-binding TS \cite{ryu,Nussinov}.

Here we consider a general class of 3D tight-binding models of fermions positioned at the vertices of a cubic lattice. We allow for tunnelling and pairing interactions between nearest and next-to-nearest neighbouring sites of the lattice. When we impose time-reversal (TR) and particle-hole (PH) symmetries we obtain explicit realisations of 3D TS of type DIII \cite{Schnyder}. For periodic boundary conditions in all three directions we provide a variety of TS that support winding numbers $\nu_\text{3D} =0, \pm1, \pm2, \pm3, \pm4$. Similar models that exhibit $\nu_\text{3D}=\pm2$ \cite{Ludwig} and $\nu_\text{3D}=\pm1$ \cite{Fu} and higher \cite{Duan} have recently been presented. In our model, the higher values of $\nu_\text{3D}$ are obtained while keeping fixed the size of the unit cell and the range of interactions. Subsequently, we impose open boundary conditions in one direction and identify edge modes with dispersion relation that crosses the band gap \cite{Hasan}, which correspond to gapless Majorana surface states. These states acquire a gap when a Zeeman field is applied at the boundary that breaks TR symmetry \cite{SatoTak,Alicea,FuKane}. Thus, the boundary behaves as a TS of the class D. It is known that both the 3D class DIII and the 2D class D TS have a $\mathbb Z$ topological invariant \cite{Schnyder}. Here we show that under certain conditions the actual values of these topological invariants that describe the bulk and the boundary physics of a 3D class DIII TS are equal. We demonstrate this both numerically, for all the tight-binding models presented here, and theoretically, based on an effective topological field theory. This protection of the boundary topological order from the topological character of the bulk provides the means for the fault-tolerant realisation of localised Majoranas \cite{Read-Green}. These Majoranas can be employed for topological quantum computation \cite{PachosB} that is resilient against erroneous perturbations or thermal fluctuations.

\begin{figure}[t]
\begin{center}
\includegraphics[scale=0.40]{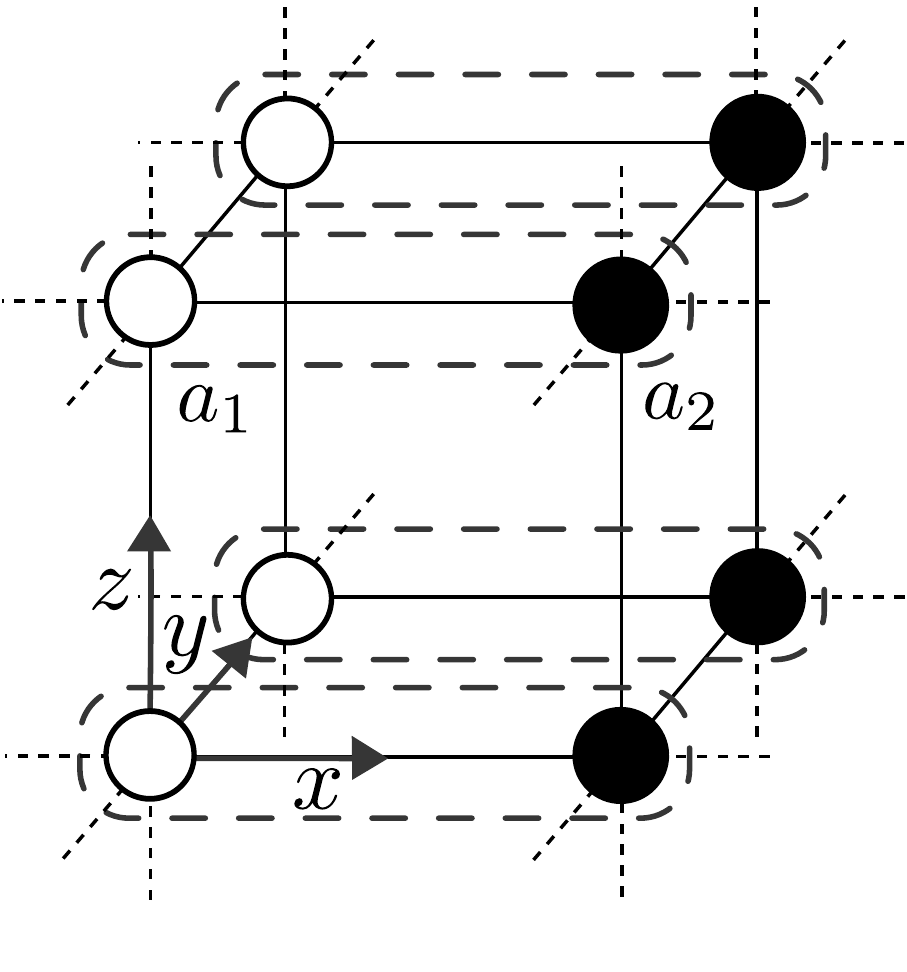}\,\,\,\,\,
\includegraphics[scale=0.25]{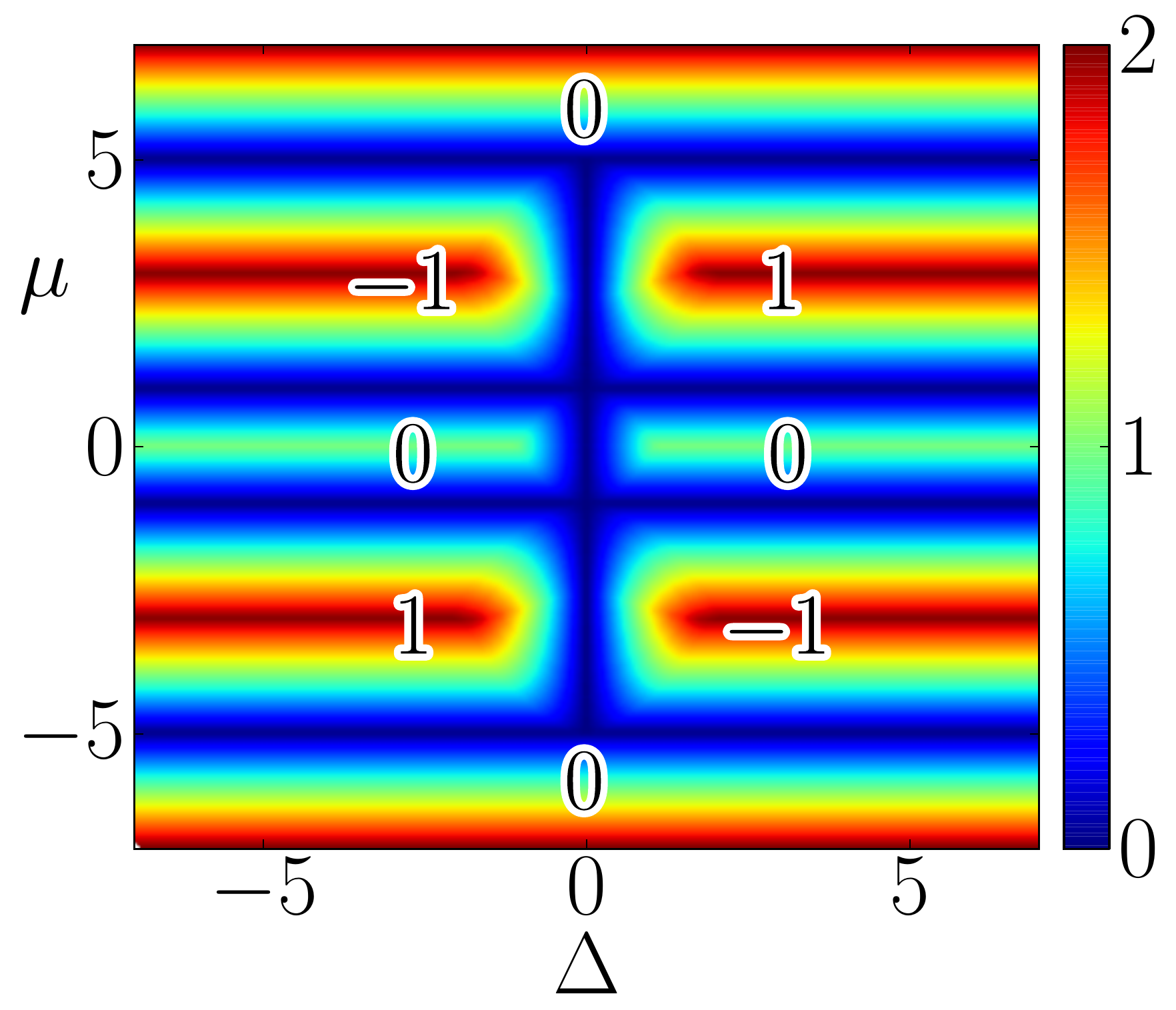}
\end{center}
\caption{\label{fig:GapWinding} (Colour online) (Left) The fermionic cubic lattice. The unit cell (dashed square) consists of two fermions $a_1$ (white site) and $a_2$ (black site). Tunnelling and pairing interactions are assumed between nearest and next-to-nearest neighbouring sites. (Right) The energy gap $\min_{\pp}(|E(\pp)|)$ as function of $\mu$ and $\Delta$ when $t=1$. The winding number is depicted as a function of $\mu$ and $\Delta$ taking values $\nu_\text{3D} = 0,\pm1$. Changes in the winding number are accompanied by quantum phase transitions.}
\end{figure}

{\bf \em DIII lattice:--} We now introduce the lattice model. We consider two species of fermion, $a_1$ and $a_2$, canonically ordered on a cubic lattice, as shown in Fig. \ref{fig:GapWinding} (Left). The unit cell, positioned at $\jj = (j_x,j_y,j_z)$, consists of two sites lying along the $x$-axis. The Hamiltonian is given by
\begin{eqnarray}\label{eqn:genham}
H = \sum_{\jj} \!\!\!\!\!\!&&\Big(\sum_k\mu a^{\dagger}_{k\jj}a_{k\jj} + \sum_{k,k',\sv}t_{kk'\!\sv}a^{\dagger}_{k\jj}a_{k'\!\jj+\sv}\nonumber\\ 
&&\,\,\,\,+\sum_{k,k',\sv} \Delta_{kk'\!\sv}a_{k\jj}a_{k'\!\jj+\sv} \Big)+ \text{h.c.},
\end{eqnarray}
where $t_{kk'\!\sv}$ and $\Delta_{kk'\!\sv}$ are the tunnelling and pairing couplings, respectively, $\mu$ is the chemical potential and $\sv$ is a vector connecting interacting unit cells. The interactions are taken to be at most between next-to-nearest neighbours. For periodic lattice we introduce the Fourier transformation $a_{k,\jj} = \sum _{\pp} e^{i\pp \cdot \jj} a_{k,\pp}$ to obtain $H = \sum_{\pp} \psi^\dagger_{\pp} h(\pp) \psi_{\pp}$, where 
$\psi_{\pp} = (a_{1,\pp},a^{\dagger}_{1,-\pp},a_{2,\pp},a^{\dagger}_{2,-\pp})^{T}$, 
$\pp\in \text{BZ}=[0,2\pi)\times[0,2\pi)\times[0,2\pi)$ and the kernel $h(\pp)$ is a $4\times 4$ hermitian matrix. To impose TR and PH symmetries we introduce the unitary operators $C_\mathrm{TR} $ and $C_\mathrm{PH}$ and demand that
\be
C_\mathrm{TR}^\dagger h^*(-\pp) C_\mathrm{TR} = h(\pp) ,\,\,\,
C_\mathrm{PH}^\dagger h^*(-\pp) C_\mathrm{PH} = -h(\pp)
\label{eqn:PH}
\ee
with $C_\mathrm{TR}^T =- C_\mathrm{TR} $ and $C^{T}_\mathrm{PH} = C_\mathrm{PH} $. These requirements guarantee that our model belongs in class DIII. Next we interpret the species index $k=1,2$ as spin-$1/2$ components. For simplicity we restrict to Hamiltonians such that, in the basis $\psi_{\pp} = (ia_{1,\pp}-a^\dagger_{1,-\pp},a_{2,\pp} - ia^\dagger_{2,-\pp},ia_{1,\pp} + a^\dagger_{1,-\pp},a_{2,\pp}+ia^\dagger_{2,-\pp})^T/\sqrt{2}$, the kernel takes the spin-triplet TS form~\cite{Ueda,Sato}
\be
h(\pp) = 
\left(
\begin{array}{ccc}
\epsilon(\pp)\mathbb{I} & \Theta(\pp)  \\
\Theta(\pp)^\dagger & -\epsilon(\pp)\mathbb{I}
\end{array}
\right),
\label{eqn:Ham2}
\ee
with $\epsilon(\pp)$ denoting the normal state, $\mathbb{I}$ is the identity $2\times2$ matrix and $\Theta(\pp)$ the spin-triplet pairing function $\Theta(\pp)=i (\boldsymbol{d}(\pp)\cdot \boldsymbol{\sigma})\sigma_{y}$. Here, $\boldsymbol{\sigma}=(\sigma_{x},\sigma_{y},\sigma_{z})$ are the Pauli matrices and $ \boldsymbol{d}(\pp)=(d_{x}(\pp),d_{y}(\pp),d_{z}(\pp))$ are odd functions. The corresponding doubly degenerate spectrum is given by $E(\pp)=\pm\sqrt{\epsilon(\pp)^{2}+|\boldsymbol{d}(\pp)|^{2}}$. We take the system to be prepared in its lowest energy, where both negative valance bands are completely occupied. The topological nature of the gapped regions is identified by the winding number $\nu_{\text{3D}}\in\mathbb{Z}$ that characterises the mapping between the toroidal Brillouin zone $T^3$ and the sphere $S^3$ defined by the normalised 4D vector $(\epsilon(\pp),\boldsymbol{d}(\pp))/|E(\pp)|$. This can be evaluated in terms of projectors to the two lowest eigenstates of kernel \eqref{eqn:Ham2}  \cite{Schnyder}. 

We now take a specific coupling configuration that gives rise to a non-trivial winding number.  More concretely, for a particular set of interactions $\{\sv\}$ we can have
$\epsilon(\pp) = t[\cos (p_x-p_z) + \cos (p_x) +2\cos (p_y-p_z)+\cos (p_y)] -\mu$, $d_{x}(\pp) = \Delta[\sin(p_x-p_z) +\sin (p_x) + 2\sin(p_y-p_z) -\sin (p_y)]$, $d_{y}(\pp) = 2\Delta \sin (p_y)$ and $d_{z}(\pp) =2\Delta\sin (p_x+p_y)$ with $t,\mu,\Delta\in\mathbb{R}$. The energy gap as function of $\mu$ and $\Delta$ for $t=1$ as well as the winding number, $\nu_\text{3D}$, corresponding to each gapped phase are shown in Fig. \ref{fig:GapWinding} (Right). It is possible to evaluate the winding number in terms of the set of momenta $\boldsymbol{p}^*$ satisfying $\boldsymbol{d}({\boldsymbol{p}}^*) =0$ from $\nu_\text{3D} = {1 \over 2}\sum_{\boldsymbol{d}(\pp^*)=0}  \text{sgn}[\epsilon (\pp^*)]\text{sgn}\{\det[\partial_jd_i(\pp^*)]\}$  \cite{Sato}. This expression shows explicitly the dependence of its sign, $\text{sgn}(\nu_\text{3D})$, in terms of the product of the signs of the couplings $\Delta$ and $\mu$, as shown in Fig. \ref{fig:GapWinding} (Right).

{\bf \em Boundary properties:--} Let us now consider the case where a boundary is introduced. For concreteness we take the lattice to extend between two disconnected planes. The Bottom plane (B) positioned at $z=1$ and the Top plane (T) positioned at $z=l$, where $l$ is a positive integer. The Hamiltonian of the system with a boundary is given by
\begin{eqnarray}
	H & = & \sum_{\bar{\pp},\zz, \sv} \psi^\dagger_{\zz,\bar{\pp}} h(\bar{\pp};\zz,\zz+\sv) \psi^{}_{\zz
	+\sv,\bar{\pp}}  \nonumber  \\
	&& + \sum_{\bar{\pp}} \left( \psi^\dagger_{1,\bar{\pp}} h_{B} \psi^{}_{1,\bar{\pp}} + \psi^\dagger_{l,\bar{\pp}} h_{T} \psi^{}_{l,\bar{\pp}} \right)
\label{eqn:bound}
\end{eqnarray}
where $\psi_{\zz,\bar{\pp}}= (a^{}_{1,\zz,\bar{\pp}},a^\dagger_{1,\zz,-\bar{\pp}}, a^{}_{2,\zz,\bar{\pp}}, a^\dagger_{2,\zz,-\bar{\pp}})^T$ with $\zz$ a vector in the $z$-direction, $\bar{\pp}\in[0,2\pi)\times[0,2\pi)$ the momentum on the $x$--$y$ plane and $h_{B}$ and $h_{T}$ are interaction terms corresponding to the Bottom and Top planes, respectively. These terms are introduced to give an energy gap to the boundary states and do not affect the properties of the bulk. While it is possible to consider independent interaction terms at each plane, for uniformity we choose them to be equal, given by $h_{T} =h_B = {\bf B}\cdot{\bf {\SIGMA}} \otimes \mathbb{I}$, where ${\bf B}$ is a 3D vector. These terms can be viewed as an effective Zeeman field. They correspond to interactions between the fermionic modes $a_1$ and $a_2$ within the same unit cell. 

\begin{figure}[t]
\begin{center}
\vspace{0.11cm}
\includegraphics[scale=0.237]{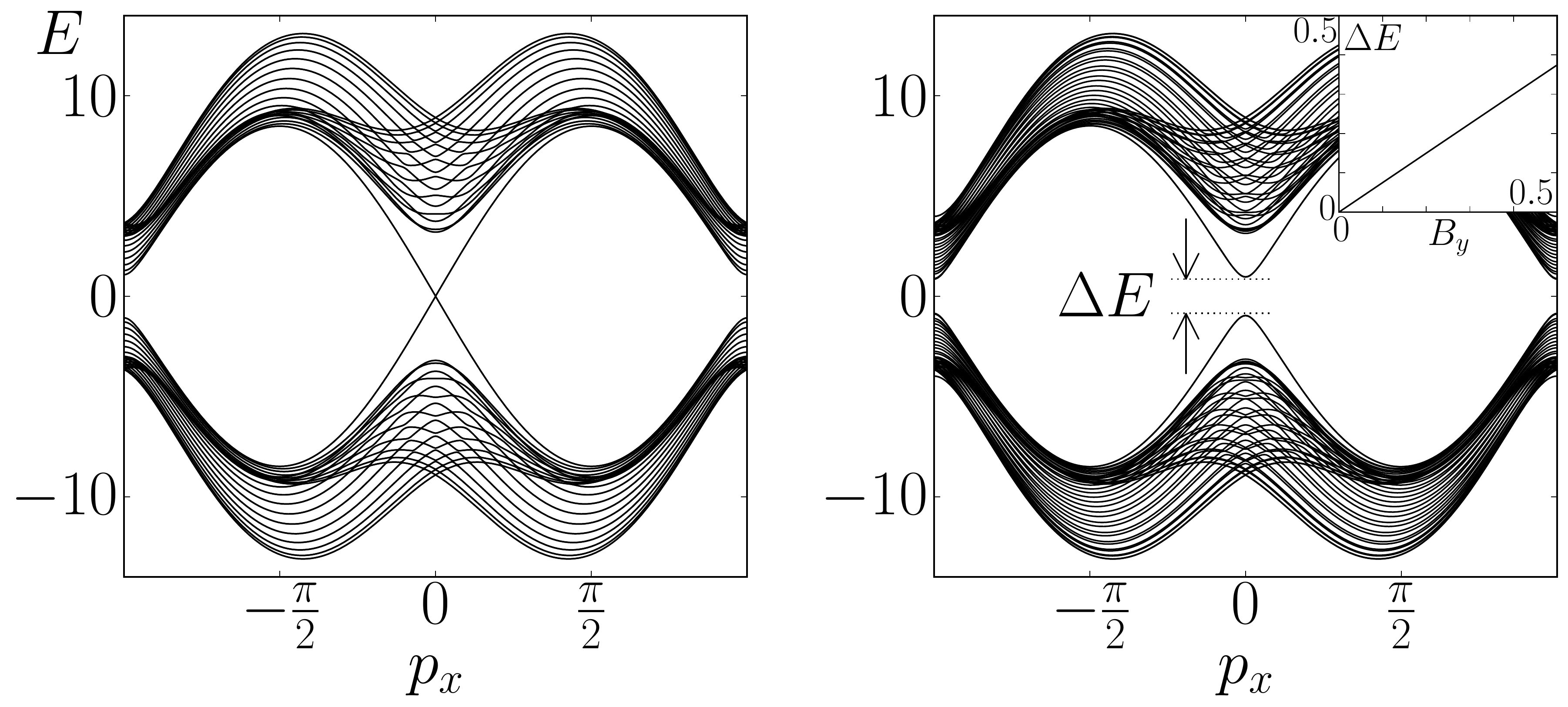}
\end{center}
\caption{\label{fig:edgestat} (Left) An energy dispersion cross section at $p_y=0$ for the $\nu_\text{3D}=1$ model ($t=1$, $\Delta=2$, $\mu=3$) with open boundaries at $z=1$ and $z=20$. A single  cone appears with a double degeneracy corresponding to the two Majorana edge modes, one for each surface. (Right) A boundary Zeeman field with $B_y\neq 0$ generates an energy gap, $\Delta E$, to the edge modes. (Inset) The energy gap $\Delta E$ as a function of $B_{y}$.}
\end{figure}

Initially, let us take the case where ${\bf B} = 0$. When the winding number $\nu_\text{3D}$ is non-zero then the bulk-boundary correspondence necessitates that gapless helical modes are present at each edge of the system \cite{Schnyder}. These are manifested as 2D gapless Majorana cones in the dispersion relation. This is verified in Fig. \ref{fig:edgestat} (Left), where the two edge modes, one per surface, appear in the spectrum as two degenerate conical states. Next, we introduce a magnetic field, ${\text{\bf B}}\neq0$, on the boundary. Non-zero values of $B_{x,z}$ change the position of the Majorana cones in momentum space. The $B_y$ term corresponds to complex tunnelling couplings so it breaks TR symmetry at the boundary. As a result it gives an energy gap to the 2D Majorana surface states that appear as paraboloids in Fig. \ref{fig:edgestat} (Right). We can evaluate the 2D winding number in terms of the projectors onto these gapped surface states~\cite{Thouless,Lisle}. For $N$ such states with the same helicity the winding number is given by $\nu_{b} = \pm N/2$ where $b=T,B$, that we call {\em partial} winding number. The sign of $\nu_{b}$ depends on the helicity of the edge states as well as the sign of $B_y$ that generates their gap.

While each plane constituting the boundary can be treated independently, the condition $h_T=h_B$ allows us to consider the entire boundary as a 2D TS dislocated between the two planes. The effective Hamiltonian that describes the low energy limit of these 2D superconducting states breaks TR symmetry, due to the presence of non-zero $B_y$, so it behaves as a class D system \cite{Schnyder}. Subsequently, we can define the sum of the partial winding numbers, $\nu_\text{2D}=\sum_{b}\nu_{b}$, which characterises the topological phase of the boundary as a whole. 

We would now like to see how the winding number $\nu_\text{2D}$ of the boundary relates to the winding number $\nu_\text{3D}$ of the bulk when both bulk and boundary are gapped. To make the comparison legitimate we choose $\Delta$, the order parameter of the bulk, and $B_y$, the order parameter of the boundary, to have the same sign, i.e. $\text{sgn}(B_y) = \text{sgn}(\Delta)$. Let us first look at the $\nu_\text{3D}=0$ case. We find that each plane supports two pairs of Majorana edge modes with {\em opposite} helicities. As a result the 2D winding number is $\nu_\text{2D}=0$. When $\nu_\text{3D}=1$ each plane supports $N=1$ Majorana cone. When  gapped, each partial winding number contributes $\nu_{b}=1/2$ to the 2D winding number, such that $\nu_\text{2D}=1$. A similar result holds when $\nu_\text{3D}=-1$. In analogy to the  2D SC in class D \cite{Read-Green} we expect that when $\nu_\text{3D}=\pm 1$ the boundaries can support Majoranas localised at the endpoints of vortex strings that terminate on the boundaries \cite{Teo}.

{\bf \em Higher winding numbers:--} We now present models that support higher winding numbers without the need to increase the size of the unit cell or the range of interactions. These models are obtained by searching among a variety of possible configurations of interactions that respect the TR and PH symmetries (\ref{eqn:PH}), so they are in class DIII. Using the notation of \eqref{eqn:Ham2}, we present Hamiltonian $H_2$ with $\epsilon(\pp) = t[\cos(p_{x})+\cos(p_{x}-p_{z})]-\mu$, $d_{x}(\pp) =\Delta[\sin(p_{x})+\sin(p_{x}-p_{z})]$, $d_{y}(\pp)=2\Delta\sin(p_{y})$ and $d_{z}(\pp)=2\Delta\sin(p_{x}+p_{y})$ that supports topological phases with $\nu_\text{3D}=0,\pm2$, Hamiltonian $H_3$ with $\epsilon(\pp) = -t[-\cos(p_{x})+\cos(p_{x}-p_{z})+\cos(2p_{y})-\cos(p_{y}+p_{z})+\cos(p_{y})]+\mu$, $d_{x}(\pp)=\Delta[-\sin(p_{x})-\sin(p_{x}-p_{z})+\sin(2p_{y})-\sin(p_{y})-\sin(p_{y}-p_{z})]$, $d_{y}(\pp)=2\Delta\sin(p_{y})$ and $d_{z}(\pp)=2\Delta\sin(p_{x}-p_{y})$ that supports $\nu_\text{3D}=0,\pm1,\pm3$ and Hamiltonian $H_4$ with  $\epsilon(\pp)=t[\cos(p_{x}+p_{z})-2\cos(p_{y}
)] +\mu$, $d_{x}(\pp)=-\Delta\sin(p_{x}+p_{z})$, $d_{y}(\pp)=-2\Delta\sin(p_{z})$ and $d_{z}(\pp)=2\Delta\sin(p_{x}+p_{y})$ that supports $\nu_\text{3D}=0,\pm2,\pm4$.

\begin{figure}[t]
\begin{tabular}{lcr}
\hspace{-0.2cm}\includegraphics[scale=0.175]{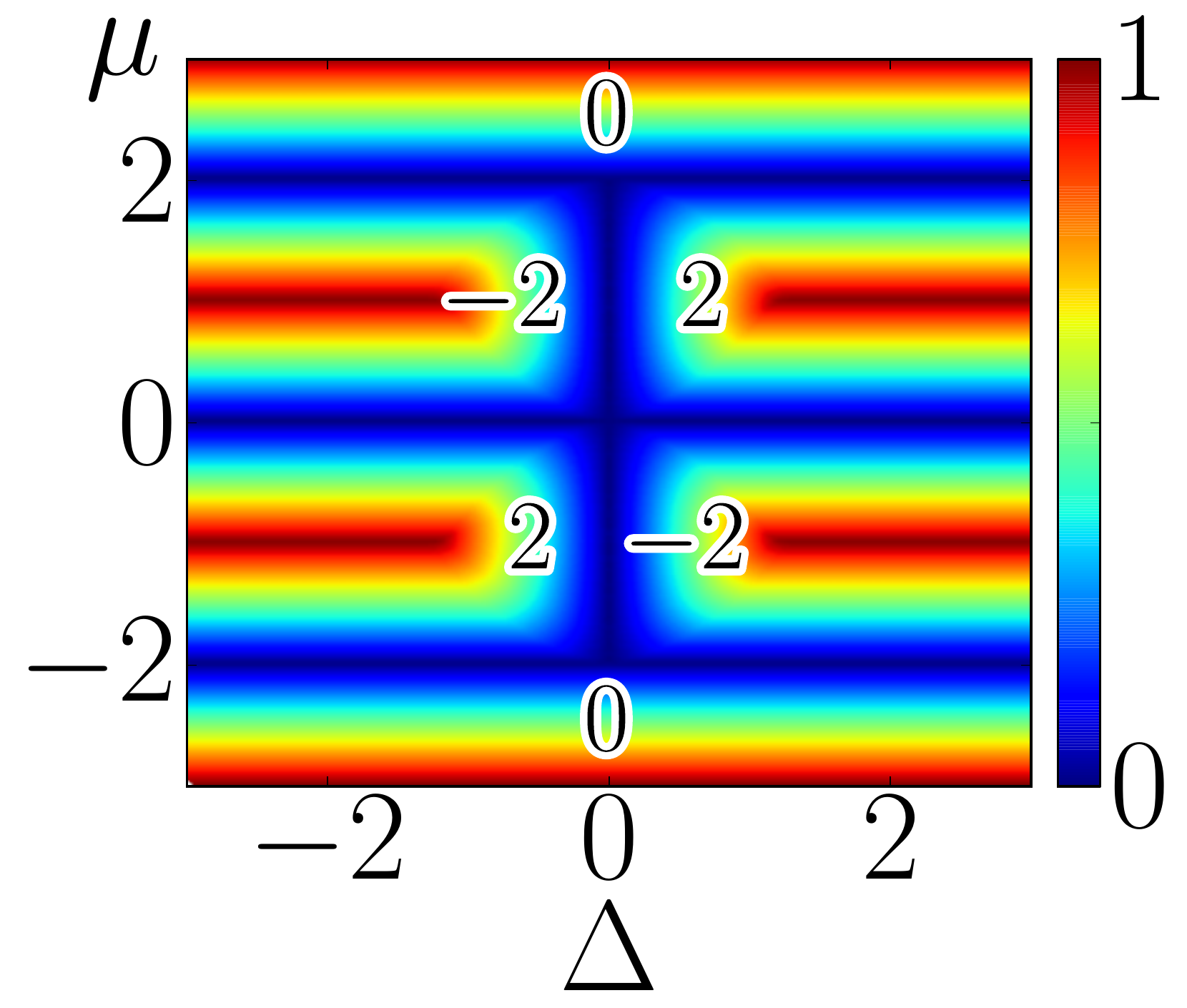}\hspace{-0.17cm}&\hspace{-0.17cm}\includegraphics[scale=0.175]{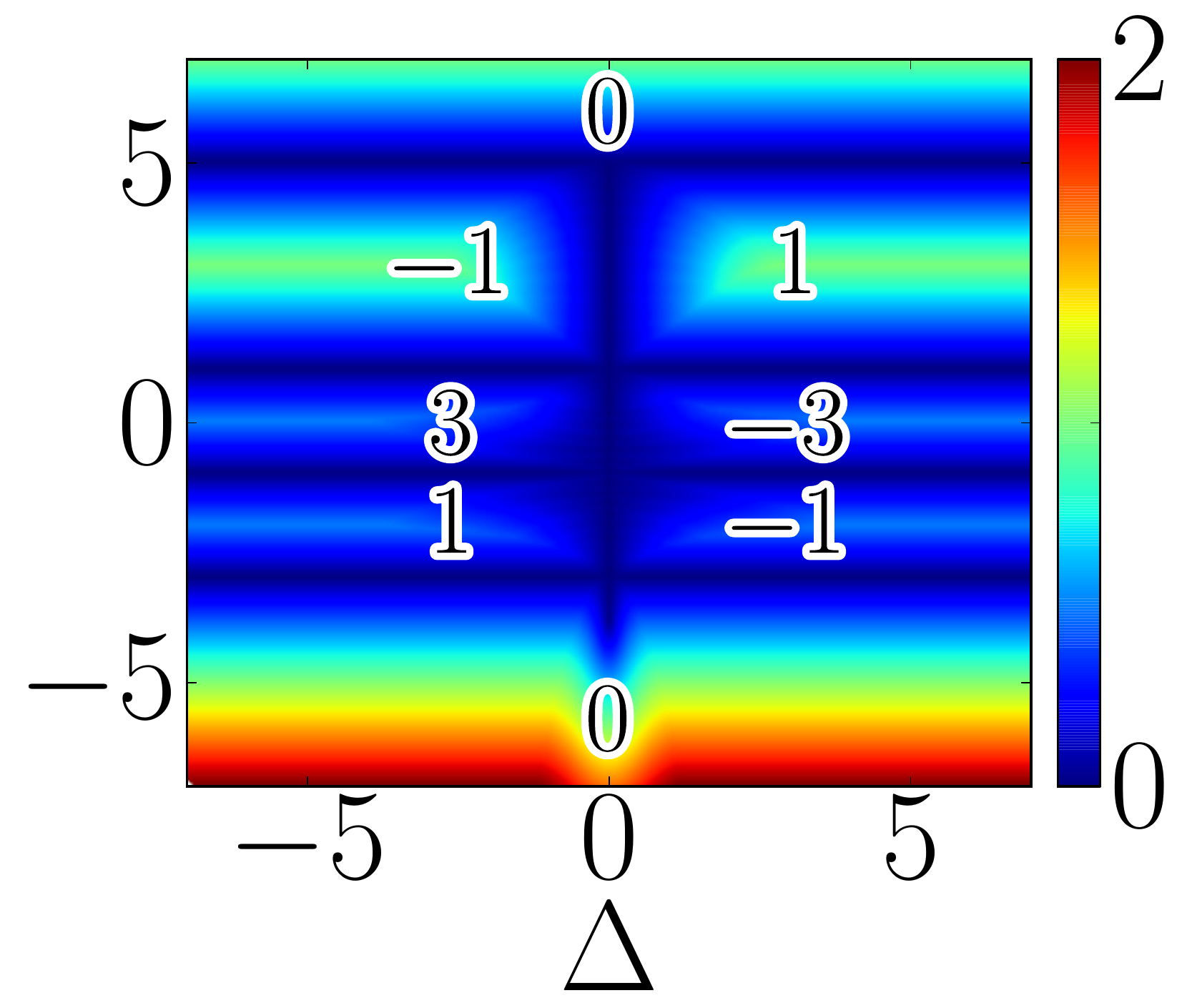}\hspace{-0.17cm}&\hspace{-0.17cm}\includegraphics[scale=0.175]{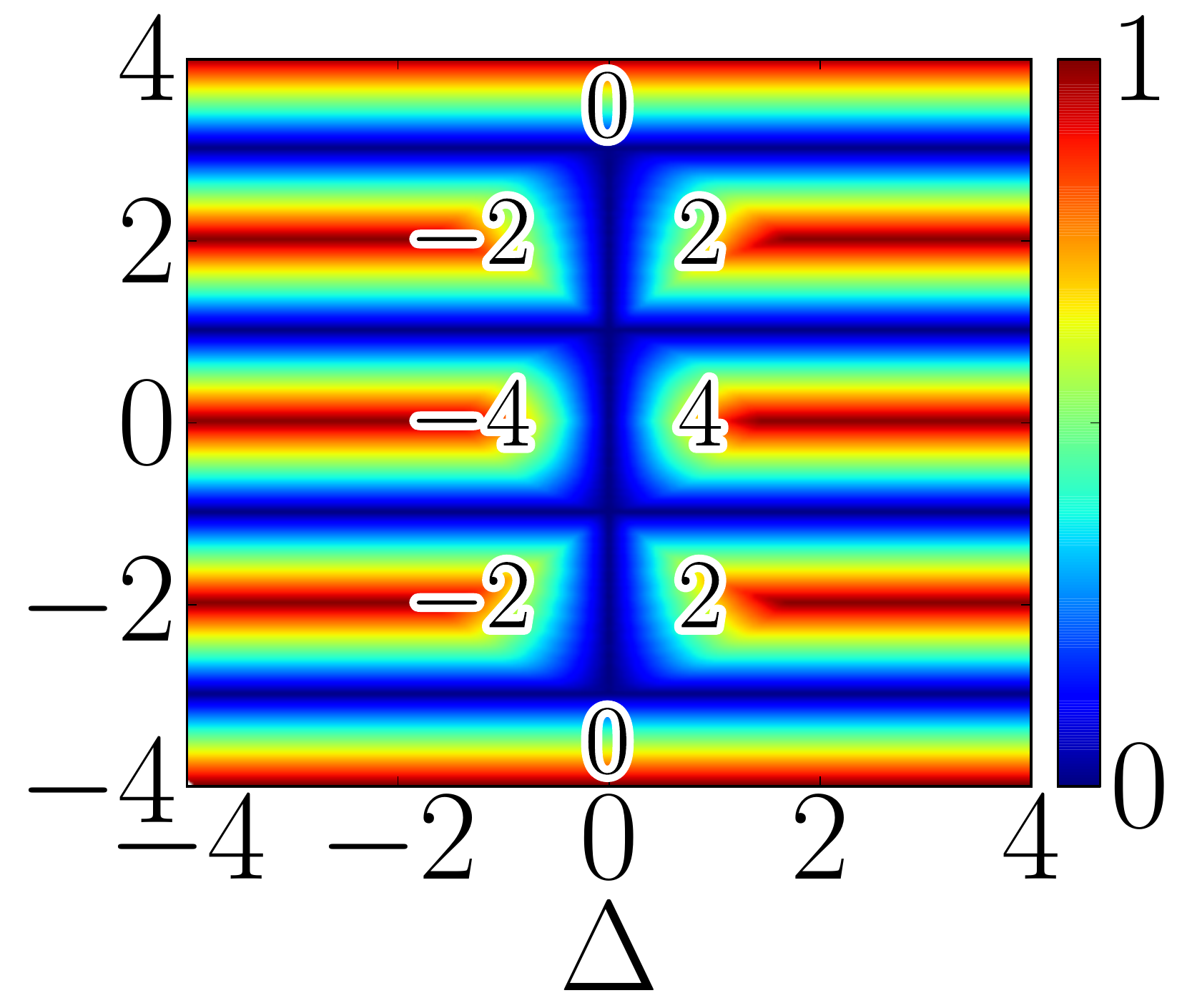}\\
\hspace{-0.18cm}\includegraphics[scale=0.175]{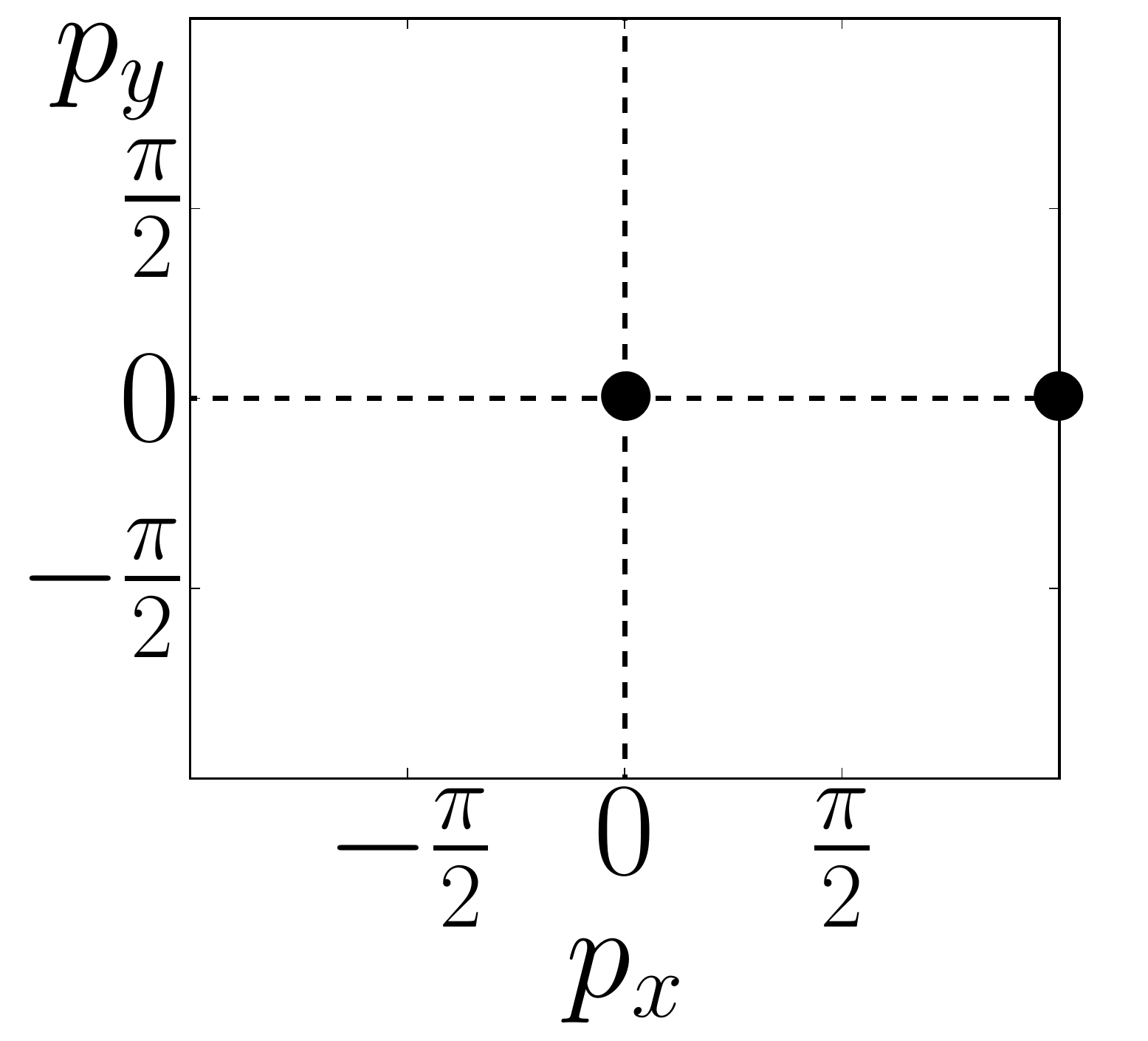}&\hspace{-0.3cm}\includegraphics[scale=0.175]{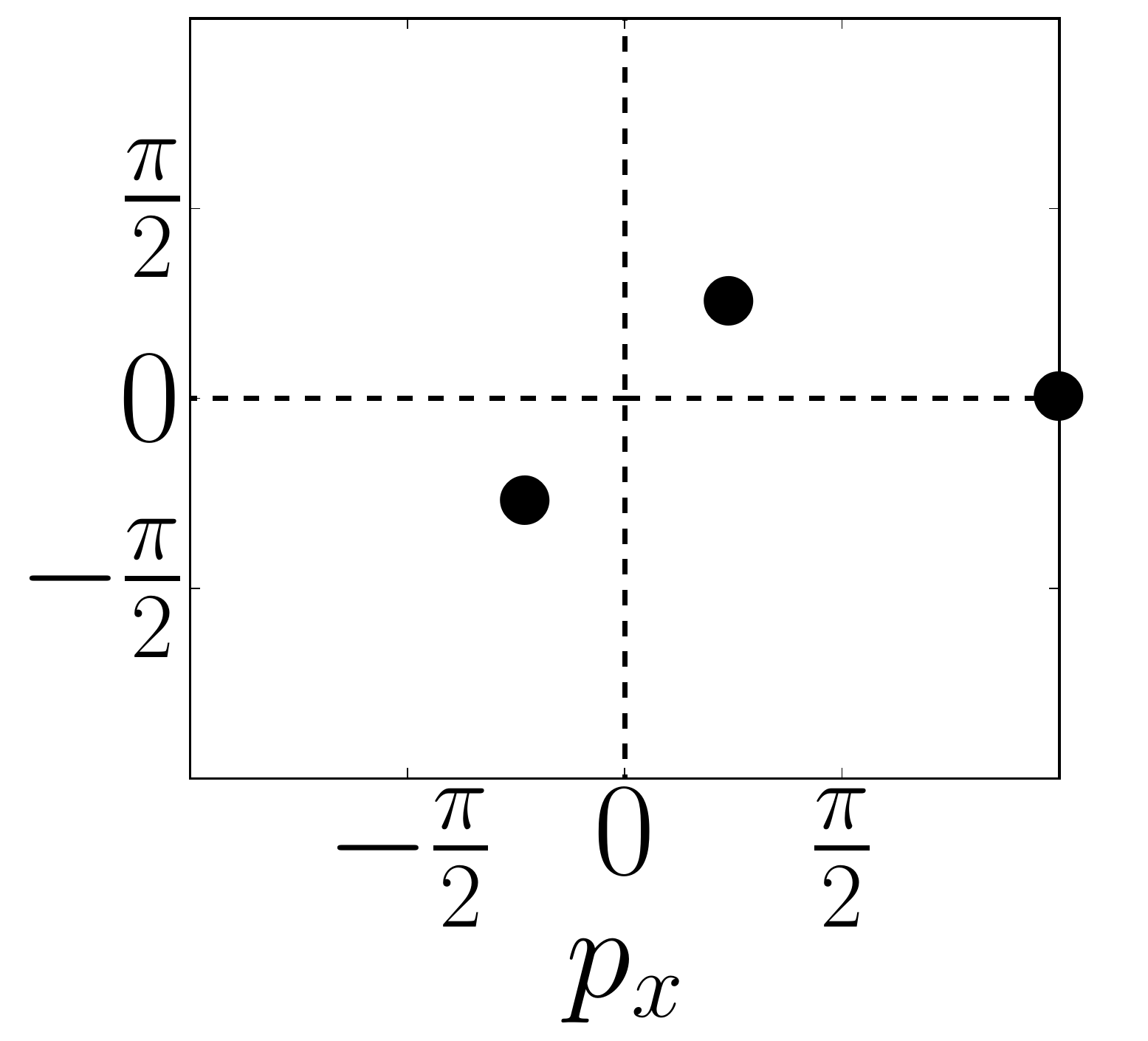}&\hspace{-0.5cm}\includegraphics[scale=0.175]{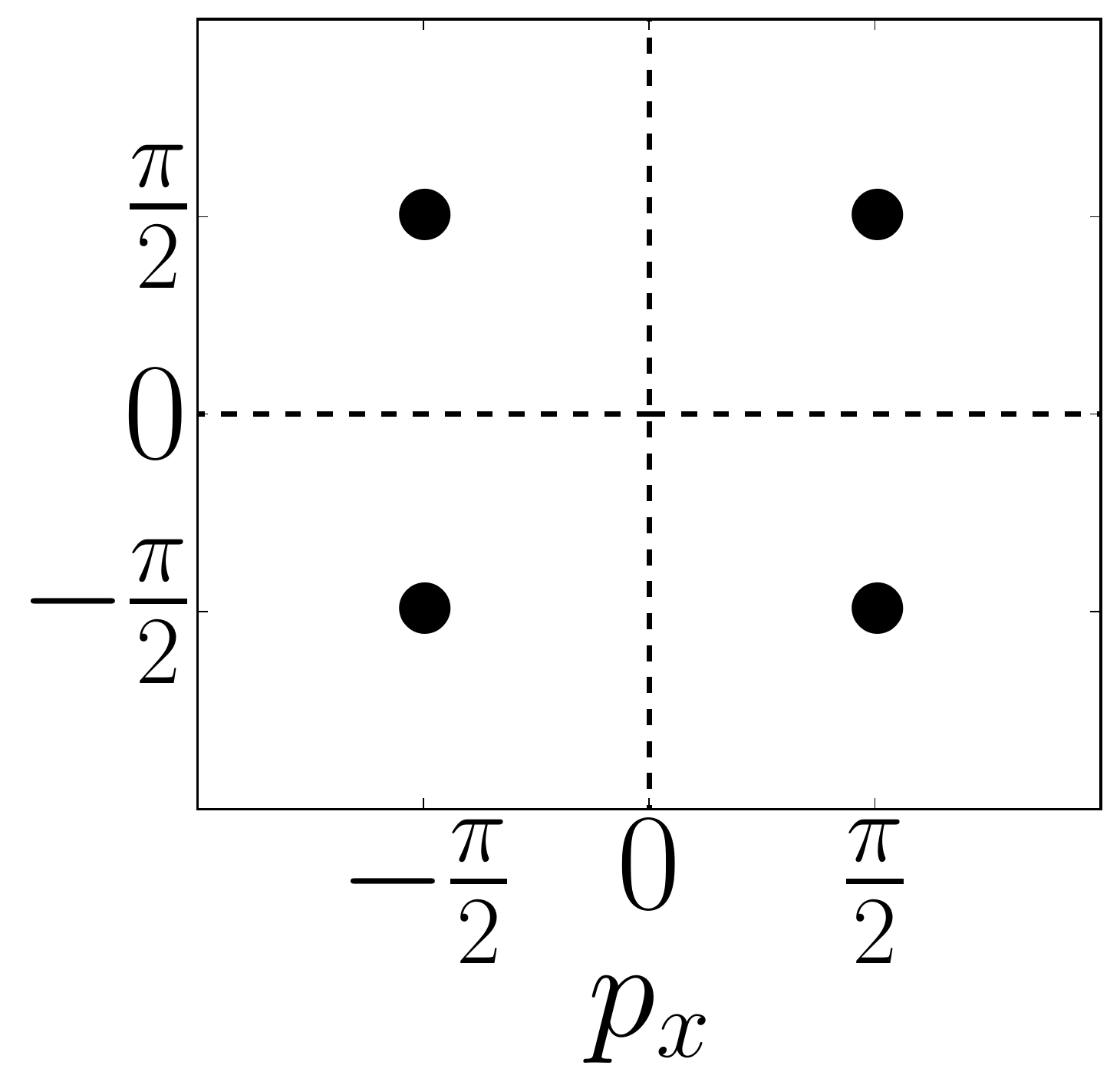}\hspace{0.35cm}
\end{tabular}
\caption{\label{fig:edgestatNEW} (Colour online) (Top Row) The energy gaps of  $H_2$, $H_3$ and $H_4$, ordered from left to right, as a function of $\mu$ and $\Delta$ for $t=1$. The winding number associated with each gapped phase is indicated. (Bottom Row) The centre of the surface Majorana cones are given for the model corresponding to the gap diagram above it, when each model is in the $\text{max}(\nu_{\text{3D}})$ phase. All plots were generated with $\mu=0.1$, $\Delta=t=1$.}
\end{figure}

The phase diagrams of the $H_2$, $H_3$ and $H_4$ models are shown in Fig. \ref{fig:edgestatNEW} (Top Row). In each case, multiple gapped topological phases are separated by gapless phase transitions. We then introduce open boundary conditions in the $z$-direction and observe gapless Majorana states localised at each surface. When the winding number $\nu_\text{3D}$ is non-zero we find that for all models $N=|\nu_\text{3D}|$ many such surface states exist at each edge with the same helicity. Next we consider each model in the phase with $\text{max}(\nu_\text{3D})$. The position of the zero energy points of their Majorana edge states is shown in Fig. \ref{fig:edgestatNEW} (Bottom Row). A pair of Majorana modes (one per edge) corresponds to each dot in the Figure. Finally, we introduce a boundary Zeeman field with $B_y\neq0$. We numerically find that for all cases the induced 2D winding number satisfies $\nu_\text{2D} = \nu_\text{3D}$. In other words, if a phase transition in the bulk occurs such that the new phase has a different 3D winding number, then the number and/or helicity of Majorana cones on the boundary and the 2D winding number change accordingly.

{\bf \em Bulk-boundary correspondence:--} The bulk-boundary correspondence witnessed here as the agreement between $\nu_\text{2D}$ and $\nu_\text{3D}$ is not accidental. We now present an analytic link between these two topological invariants by looking at the thermal Hall conductivity (THC). In particular, we establish a relationship between the THC of a 2D TS in the class D, that is proportional to $\nu_\text{2D}$, and the THC induced on the boundary (composite system of two surfaces) of a 3D TS in the class DIII, that is shown to be proportional to $\nu_\text{3D}$.

The thermal properties of TS can be encoded in an effective topological field theory (ETFT) obtained by introducing a fictitious gravitational field, as described by Luttinger theory \cite{Luttinger}. In the 2D class D model, the ETFT is a gravitational Chern-Simons theory as showed in \cite{Read-Green, Volovik} (see also \eqref{eqn:CS} below). From this description and from independent arguments \cite{Fujimoto2D} it was possible to show that the THC is given by $\kappa_{\text{th}}^{\text{2D}}=\frac{\nu_{C}}{12}\pi \,T$, for temperatures $T\rightarrow 0$, where $\nu_{C}$ is the Chern number of the TS ($\hslash=k_{B}=1$). In our case, the THC of the composite boundary of the 3D TS is the sum of the THC of the two surfaces, thus we have
\be
\kappa_{\text{th}}^{\text{boundary}}=\frac{\nu_\text{2D}}{12}\pi \,T,
\label{thermH2d}
\ee
where $\nu_\text{2D}=\nu_{T}+\nu_{B}$.

We now calculate the boundary THC in an alternative way, by starting from the 3D bulk properties. The 3D model (\ref{eqn:genham}) of a certain topological phase can be adiabatically connected to a model with a low energy description given by massive Dirac fermions \cite{Ryu2} with action
\be
S_{\psi}=\int_{M} d^{4}x \,\bar{\psi}\big(\gamma^{\mu}\partial_{\mu}+m\big)\psi,
\label{eqn:action1}
\ee
where $\mu=0,1,2,3$, $\bar{\psi}=\gamma^{0}\psi$, $\gamma^{\mu}$ are the Dirac matrices and $M$ is the $(3+1)$-dimensional spacetime. In order to calculate the ETFT, we introduce a curved background to the fermionic action $S_{\psi}$. This is done by coupling the fermions with spin connection, $\omega_{\mu}$, and tetrads that naturally encode the geometric properties of curved spaces \cite{Note}. The effective action $S_{\text{eff}}$ that describes the model purely in terms of the spin connection can be derived by integrating out the fermions in the corresponding partition function. The topological part of $S_{\text{eff}}$ that faithfully describes the low-energy behaviour of the model is given by \cite{Qi-Zhang, Ryu-Moore}
\be
S_{\text{eff}}^{M,\text{top}}=\frac{1}{2}\frac{\theta}{768 \pi^{2}}\int_{M} d^{4}x\, \epsilon^{\mu\nu\alpha\beta}\text{tr}(R^{\rho}_{\sigma\mu\nu}R^{\sigma}_{\rho\alpha\beta}),
\label{eqn:topact}
\ee
where $R^{\rho}_{\sigma\mu\nu}$ is the Riemann tensor given in terms of the spin connection $\omega_{\mu}$, while $\theta=\nu_\text{3D} \pi$ mod $2\pi$ \cite{Wang-Zhang}. This is the ETFT of the 3D TS. The topological behaviour of the gapped boundary can be obtained from \eqref{eqn:topact} by considering that the spatial part of $M$ has a non-empty boundary with spacetime boundary manifold given by $\partial M$. By applying Stokes' theorem we have
\be
S_{\text{eff}}^{\partial M,\text{top}}=\frac{1}{2}\frac{\theta}{192 \pi^{2}}\int_{\partial M} d^{3}x\, \epsilon^{\mu\nu\lambda}\text{tr}
\Big(\omega_{\mu}\partial_{\nu}\omega_{\lambda}+\frac{2}{3}\omega_{\mu}\omega_{\nu}\omega_{\lambda}
\Big).
\label{eqn:CS}
\ee
The action $S_{\text{eff}}^{\partial M,\text{top}}$ corresponds to the gravitational Chern-Simons theory. By general arguments \cite{Qi-Zhang,Ryu-Moore,Read-Green} one can connect the coefficient of this Chern-Simons theory to the THC on the boundary of a 3D TS. In our system the boundary consists of two disconnected surfaces, the Top and the Bottom. Therefore the total boundary THC, $\kappa_{\text{th}}^{\text{boundary}}$, is given by
\be
\kappa_{\text{th}}^{\text{boundary}}=2\times\frac{\nu_\text{3D}}{24}\pi \,T.
\label{eqn:thermH}
\ee
Comparing equations \eqref{thermH2d} and \eqref{eqn:thermH}, we deduce that 
\be
\nu_\text{3D}=\nu_\text{2D}
\label{eqn:relation}
\ee
as it was also numerically verified for all the presented models. Note that a relative sign freedom in (\ref{eqn:relation}) is possible due to the orientation of the boundary surface employed during the application of the Stokes' theorem. This is the geometric equivalent of the freedom we had in choosing the sign of $B_y$.

{\bf \em Protected 2D topological order:--} We have presented 3D TS models in class DIII that exhibit a large variety of winding numbers $\nu_\text{3D} = 0,\pm 1,\pm 2, \pm 3, \pm 4$. Numerical and theoretical analysis showed that the boundary of these models behaves as a 2D TS of class D with winding number $\nu_\text{2D} = \nu_\text{3D}$. Nevertheless, there is an intriguing difference between the boundary of a 3D TS and a purely 2D TS system. In the 3D case the 2D topological boundary is {\em delocalised} between two spatially separated surfaces. This non-local encoding of topological order, together with relation (\ref{eqn:relation}), provide a protection that is not present in the purely 2D system, as we analyse below. 

Firstly, external perturbations cannot affect the topological nature of the boundaries. As an adversary mechanism consider placing on a $\nu_b=1/2$ edge another 2D lattice system with Chern number $\nu_\text{Ch} = \pm1$. The latter can be effectively described by two massive Majorana fermions~\cite{KitaevHoney}. Perturbative interactions between this system and the edge can cause the edge Majorana fermion to pair with one of the Majorana fermions of the appended system creating a non-topological Dirac fermion \cite{Bernevig}. This leaves behind an edge with a single Majorana fermion that is still described by $\nu_b=\pm1/2$, where one can easily compensate for a change in the sign. In particular, this mechanism cannot cause $\nu_b$ to become zero. 

Secondly, phase transitions due to a thermal environment are suppressed at low enough temperatures. Let us consider local thermal fluctuation at the boundary of the TS manifested as a vortex string in the bulk with both endpoints (a vortex-antivortex pair) residing on the same surface. As these endpoints bear localised Majoranas, a finite density of such thermal errors could cause a quantum phase transition due to vortex nucleation \cite{Ville,Bauer}. Nevertheless, these vortex strings have a finite energy per string length that generates a string tension \cite{Eto}. As a consequence, the vortex and the antivortex will be attracted to each other and annihilate, thus self-correcting the thermal error \cite{Schakel}. Only vortex strings with end-points at opposite surfaces can be stable \cite{Hosur,Hung}. Hence, for low enough temperatures, the vortex string tension will cause the density of vortices on the surface to be zero, and its topological phase will remain intact. These error-resilience characteristics make the boundaries of 3D class DIII TS a promising laboratory for performing topological quantum computation \cite{PachosB}. 

{\bf \em Acknowledgements:--} We would like to thank Ville Lahtinen, Joel Moore and Steven Simon for inspiring discussions. This work was supported by EPSRC.


\begin{thebibliography}{25}
\bibitem{Note}
We apply the following substitutions in the action (\ref{eqn:action1}): $\partial_{\mu} \rightarrow\partial_{\mu}+\omega_{\mu}$, 
$\gamma^{\mu}\rightarrow\gamma^{\alpha}e_{\alpha}^{\mu}$, $d^{4}x\rightarrow d^{4}x\, |e|$ with $\alpha,\beta=0,1,2,3$, where
$\omega_{\mu}=\frac{i}{4}\omega_{\mu}^{\alpha \beta}[\gamma_{\alpha},\gamma_{\beta}]$ is the spin connection, $e_{\alpha}^{\mu}$ are the tetrads and $|e|$ is the corresponding determinant. In particular, we have that $g_{\mu\nu}=e^{\alpha}_{\mu}e^{\beta}_{\nu}\eta_{\alpha\beta}$  where $g_{\mu\nu}$ is the metric tensor and $\eta_{\alpha\beta}$ the Minkowski one, while the spin connection has the Levi-Civita form: $\omega_{\mu}^{\alpha \beta}=e^{\alpha}_{\nu}\triangledown_{\mu} e^{\beta\nu}$, where $\triangledown_{\mu}=e_{\mu}^{\lambda}(\partial_{\lambda}+\omega_{\lambda})$.
\end{thebibliography}
\end{document}